\documentclass[%
reprint,
superscriptaddress,
 amsmath,amssymb,
 aps,
pre
]{revtex4-2}

\usepackage{graphicx}
\usepackage{dcolumn}
\usepackage{bm}

\begin{document}

\preprint{APS/123-QED}

\title{High-energy transient gas pinholes via saturated absorption}

\author{K. Ou}
\email{ouke025@stanford.edu}
\author{V. M. Perez-Ramirez}
\author{S. Cao}
\author{C. Redshaw}
\author{J. Lee}
\affiliation{Department of Mechanical Engineering, Stanford University, Stanford, California 94305}

\author{M. M. Wang}
\affiliation{Department of Electrical and Computer Engineering, Princeton University, Princeton, New Jersey 08540}
\author{J. M. Mikhailova}
\affiliation{Department of Mechanical and Aerospace Engineering, Princeton University, Princeton, New Jersey 08540}
\author{P. Michel}
\affiliation{Lawrence Livermore National Laboratory, Livermore, California 94551}
\author{M. R. Edwards}
\email{mredwards@stanford.edu}
\affiliation{Department of Mechanical Engineering, Stanford University, Stanford, California 94305}

\date{\today}
             
\begin{abstract}
This letter presents a spatial filter based on saturated absorption in gas as a replacement for the solid pinhole in a lens-pinhole-lens filtering system. We show that an ultraviolet laser pulse focused through ozone will have its spatial profile cleaned if its peak fluence rises above the ozone saturation fluence. Specifically, we demonstrate that a 5 ns 266 nm beam with 4.2 mJ of initial energy can be effectively cleaned by focusing through a 1.4\% ozone-oxygen mixture, with about 76\% of the main beam energy transmitted and 89\% of the side lobe energy absorbed. This process can be adapted to other gases and laser wavelengths, providing alignment-insensitive and damage-resistant pinholes for high-repetition-rate high-energy lasers.
\end{abstract}

\maketitle
Substantial efforts have been made to control and improve the spatial profiles of high-energy lasers \cite{auerbach1994serrated, hunt1977improved, potemkin2007spatial, boyd2009self, Murray:00}, since the focusability of a beam depends strongly on its spatial quality. Consequently, spatial filters play a critical role in modern laser systems, used to remove high-frequency spatial noise and clean beam profiles. Filters also reduce diffraction ripples and suppress noise from parasitic lasing, amplified spontaneous emission, and self-focusing \cite{hunt1977improved, potemkin2007spatial, boyd2009self}. A typical spatial filter consists of a lens pair and a solid pinhole, where the latter removes high-frequency components of a beam by physically blocking them at the focal plane. However, this approach becomes challenging to implement when the laser is intense enough to damage the pinhole. In addition, conventional filters require precise alignment and can be vulnerable to vibrations or pointing instability.

In addition to solid pinholes, nonlinear media have been suggested for spatial filtering because the side lobes of a beam have lower intensity than the main peak, so an intensity-dependent response at the focal plane can be used to remove high-frequency components. Nonlinear mechanisms for profile cleaning include harmonic generation \cite{szatmari1997active}, ionization \cite{edwards2020multi}, and light-induced molecular orientation \cite{kato1996nonlinear}. An alternative approach is to use saturated absorption of light in nonlinear materials, where absorption drops significantly when beam fluence rises above the saturation threshold. Previous studies have successfully used dye cells to clean lasers with intensity up to $10^{10}\,\mathrm{W/cm^2}$ \cite{penzkofer1979apodizing, sinha2006saturable}.

\begin{figure}[tb]
\centering
    \includegraphics[width=\linewidth]{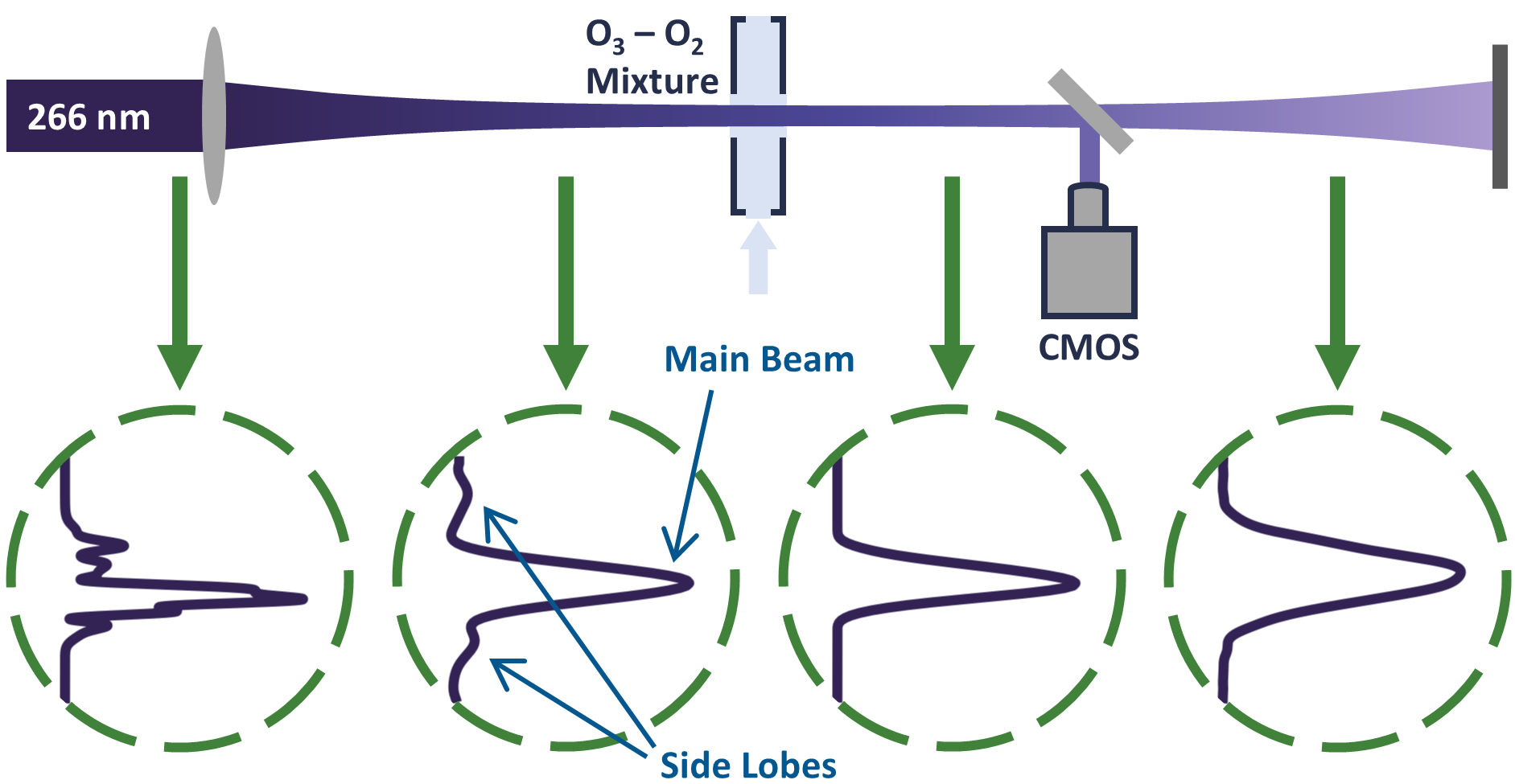}
    \caption{Experimental schematic. An aberrated $266\,\mathrm{nm}$ beam is focused into ozone. By carefully choosing the focal length, tube width, and ozone concentration, the spatially separated high-frequency components at the focal plane are mostly absorbed while the main beam is only mildly attenuated due to saturated absorption, resulting in a cleaner beam profile.}
    \label{fig:fig_schematic}
\end{figure}

In this work, we propose the use of gas as a saturated absorber for spatial filtering. Gases, compared to solid or liquid materials, have orders-of-magnitudes-higher damage thresholds. Recent work has demonstrated the creation of optics like diffraction gratings in gases, with effective damage thresholds expected to be $\sim1\,\mathrm{kJ/cm^2}$ \cite{michine2020ultra, michel2024photochemically}. Here, we demonstrate the ability of ozone to clean ultraviolet (UV) beams via simulations and experiments. As shown in Fig.\,\ref{fig:fig_schematic}, a $266\,\mathrm{nm}$ laser focused into ozone will exhibit a substantially cleaner profile after exiting the flow. Compared to other spatial filtering techniques, a gas pinhole is relatively straightforward to build, does not require precise alignment, and constantly renews itself. The last feature is a unique advantage, making this design particularly useful for high-energy lasers running at a high repetition rate.

\begin{figure*}[tb]
    \centering
    \includegraphics[width=\linewidth]{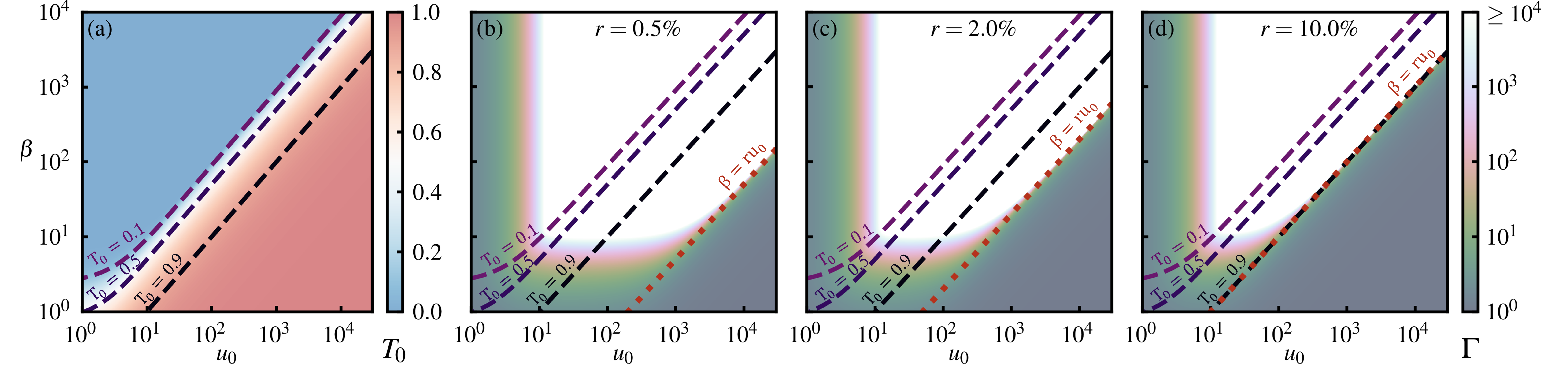}
    \caption{Transmittance $T_0$ (a) and noise attenuation $\Gamma$ (b)-(d) with various normalized fluence $u_0$ and triple product $\beta$ at $r = 0.5\%$, $2\%$ and $10\%$. The dashed contours indicate where $T_0$ reaches $0.1$, $0.5$, and $0.9$. The dotted contours indicate where $\beta = ru_0$.}
    \vspace{-10pt}
    \label{fig:fig_params}
\end{figure*}

The propagation of a pulsed laser in a saturable absorber can be described by the equations \cite{michel2024photochemically, siegman1986lasers}:
\begin{equation}
    \frac{\partial I(z,\,t)}{\partial z} = -\sigma n(z,\,t)I(z,\,t)
    \label{eq:eq1}
\end{equation}
\begin{equation}
    \frac{\partial n(z,\,t)}{\partial t} = -\frac{\sigma}{\hbar\omega} n(z,\,t)I(z,\,t)
    \label{eq:eq2}
\end{equation}
Here, $I$ is the laser intensity, $n$ is the number density of saturable absorber molecules, $\sigma$ is the absorption cross section, $\hbar$ is the reduced Planck's constant, $\omega$ is the angular frequency of the light, and $z$ and $t$ are the position and time in a reference frame that moves with the pulse (i.e. $z = \hat{z},\,t = \hat{t} - \hat{z}/c$ where $\hat{z}$ and $\hat{t}$ are the laboratory frame position and time, and $c$ is the speed of light). To find the density of the absorbing gas species after the pulse propagation, we can integrate \eqref{eq:eq2} over the pulse duration $\tau$: 
\begin{equation}
    n(z,\,t>\tau) = n(z,\,0)\exp{\left
    [-\frac{\sigma}{\hbar\omega}\int_0^\tau I(z,\,t) dt\right]} = n_0\mathrm{e}^{-U/U_s}
    \label{eq:eq3}
\end{equation}
where we assume an initially uniform absorber density $n_0$, $U$ is the beam fluence, and $U_s \equiv \hbar\omega/\sigma$ is the saturation fluence. Substituting \eqref{eq:eq2} into \eqref{eq:eq1} and integrating over the pulse duration, we obtain:
\begin{equation}
    \frac{du}{dz} = \sigma n_0(\mathrm{e}^{-u} - 1)
    \label{eq:eq4}
\end{equation}
where $u(z) \equiv U(z)/U_s$ is the normalized fluence. Finally, integrating \eqref{eq:eq4}, we get:
\begin{equation}
    \frac{\mathrm{e}^{u(z)} - 1}{\mathrm{e}^{u(0)} - 1} = \mathrm{e}^{-\sigma n_0z}
    \label{eq:eq5}
\end{equation}
If the peak and side lobes of a beam at focus have initial fluences $u_0$ and $u_1$, respectively, the transmittance $T = u(L) / u(0)$ of the two components, after propagation through a gas pinhole of length $L$ will, according to \eqref{eq:eq5}, satisfy:
\begin{equation}
    \frac{\mathrm{e}^{T_0u_0} - 1}{\mathrm{e}^{u_0} - 1} = \frac{\mathrm{e}^{T_1u_1} - 1}{\mathrm{e}^{u_1} - 1} = \mathrm{e}^{-\beta}
    \label{eq:eq6}
\end{equation}
with $\beta \equiv \sigma n_0L$ defined as the triple product of the gas pinhole. Equation \ref{eq:eq6} implies that changing the absorption cross-section, the number density, or the length of the gas pinhole are equivalent. Let $u_1 = ru_0$ where $r$ is analogous to the inverse of a signal-to-noise ratio that quantifies the noisiness of a beam. In practice, we can estimate $r$ by taking the Fourier transform of the beam profile and examining the ratio of the central mode amplitude to those of the higher-order modes. Defining the noise attenuation ratio as $\Gamma \equiv T_0/T_1$, Equation \ref{eq:eq6} can then be written as:
\begin{equation}
    \frac{\mathrm{e}^{T_0u_0} - 1}{\mathrm{e}^{u_0} - 1} = \frac{\mathrm{e}^{\frac{rT_0}{\Gamma}u_0} - 1}{\mathrm{e}^{ru_0} - 1} = \mathrm{e}^{-\beta}
    \label{eq:eq7}
\end{equation}
For an ideal gas pinhole, we would like the main beam to be entirely transmitted (i.e. $T_0 \rightarrow 1$) and the side lobes to be completely suppressed ($T_1 \rightarrow 0$ or $\Gamma \rightarrow \infty$). Given properties of the beam and gas pinhole $u_0$, $r$, and $\beta$, we can solve for $T_0$ and $\Gamma$ from \eqref{eq:eq7}:
\begin{equation}
    T_0 = \frac{-\beta + \ln{\left(\mathrm{e}^{\beta} + \mathrm{e}^{u_0} - 1\right)}}{u_0}
    \label{eq:eq8}
\end{equation}
\begin{equation}
    \Gamma = \frac{r\left[-\beta + \ln{\left(\mathrm{e}^{\beta} + \mathrm{e}^{u_0} - 1\right)}\right]}{-\beta + \ln{\left(\mathrm{e}^{\beta} + \mathrm{e}^{ru_0} - 1\right)}}
    \label{eq:eq9}
\end{equation}
Figure \ref{fig:fig_params} shows the transmittance $T_0$ and noise attenuation ratio $\Gamma$ of a gas pinhole for various initial fluence $u_0$ and triple product $\beta$ at $r = 0.5\%$, $2\%$, and $10\%$. Unlike traditional spatial filters, the transmittance of a gas pinhole is fluence-dependent. Attenuation is also limited by the beam noisiness, $r$. As $r \rightarrow 1$, $\Gamma \rightarrow 1$, indicating that spatial cleaning with a gas pinhole becomes impossible. An effective gas pinhole ($T_0\approx1$, $\Gamma \gg 1$) requires $u_0>\beta$ for good transmission, $\beta > r u_0$ for good noise attenuation, and peak fluences well above the saturation fluence, leading to a useful operating regime where
\begin{equation}
u_0 > \beta > r u_0 \gg 1
\label{eq:eq10}
\end{equation}
This condition generally gives $e^{u_0} \gg e^{\beta} \gg e^{ru_0} \gg 1$, allowing simplification of \eqref{eq:eq8} and \eqref{eq:eq9} to $T_0 \approx 1 - r - \Delta \beta / u_0$ and $\Gamma \approx r u_0 T_0 e^{\Delta \beta}$, where $\Delta \beta = \beta - ru_0 > 0$. There is therefore a trade-off between high $T_0$ and high $\Gamma$. The minimum $u_0$ depends on the the maximum achievable waist size while the maximum $u_0$ is limited by how tightly a beam can be focused and the nonlinear threshold of the gas. The feasibility of a gas pinhole depends on whether a triple product that satisfies \eqref{eq:eq10} can be produced.

We first validated this process with numerical simulations in which a beam propagates through an absorptive medium modeled by an fluence-dependent imaginary refractive index. The simulation also captures the change of the fluence as a beam propagates. Propagation of a paraxial wave in matter is governed by:
\begin{equation}
    \left.\Big[\nabla^2_\perp + 2ik_0\partial_z\right.\Big]\mathbf{E}(\mathbf{r}) = -2k_0^2\,\mathbf{E}(\mathbf{r})\frac{\delta n(\mathbf{r})}{n_0}
    \label{eq:eq11}
\end{equation}
where $\mathbf{E}$ is the electric field, $k_0$ is the vacuum wavenumber, and $n_0$ and $\delta n$ are the background and change of the refractive index. The wave equation is solved numerically using a Strang splitting scheme, with each step along $z$ combining the solutions from the paraxial propagation in vacuum via a Fourier transform along $x$ and $y$ \cite{michel2023introduction} and the absorption step, leading to:
\begin{multline}
    \hspace{-10pt}
    E(x,y,z_{j+1}) = \mathcal{F}^{-1}\left\{\mathcal{F}\left[E(x,y,z_{j})\right]\exp{\left(-\frac{ik_{\perp}^2}{2k_0}\Delta z\right)}\right\}\times\\\exp{\left[ik_0\frac{\delta n(x,y,z_j)}{n_0}\Delta z\right]}
    \label{eq:eq12}
\end{multline}
where $\mathcal{F}$ indicates Fourier transform, $k_\perp$ is the transverse wavenumber, $\Delta z$ is the propagation step size, and $j$ is the step number. Without loss of generality, we can assume that $E$ is normalized such that $|E^2| = u = U/U_s$. In addition, the absorption step is derived from discretizing \eqref{eq:eq4}:
\begin{equation}
    u_{j+1} = u_j\left[1 + \frac{\exp{(-u_j) - 1}}{u_j}\,\sigma n_0\Delta z\right].
    \label{eq:eq13}
\end{equation}
Approximating the absorption term in \eqref{eq:eq12} with \eqref{eq:eq13} gives:
\begin{equation}
    \exp{\left[ik\frac{\delta n(x,y,z_j)}{n_0}\Delta z\right]} = \sqrt{1 + \frac{\exp{(-|E_j^2|) - 1}}{|E_j^2|}\,\sigma n_0\Delta z}.
    \label{eq:eq14}
\end{equation}

\begin{figure}[tb]
    \centering
    \includegraphics[width=\linewidth]{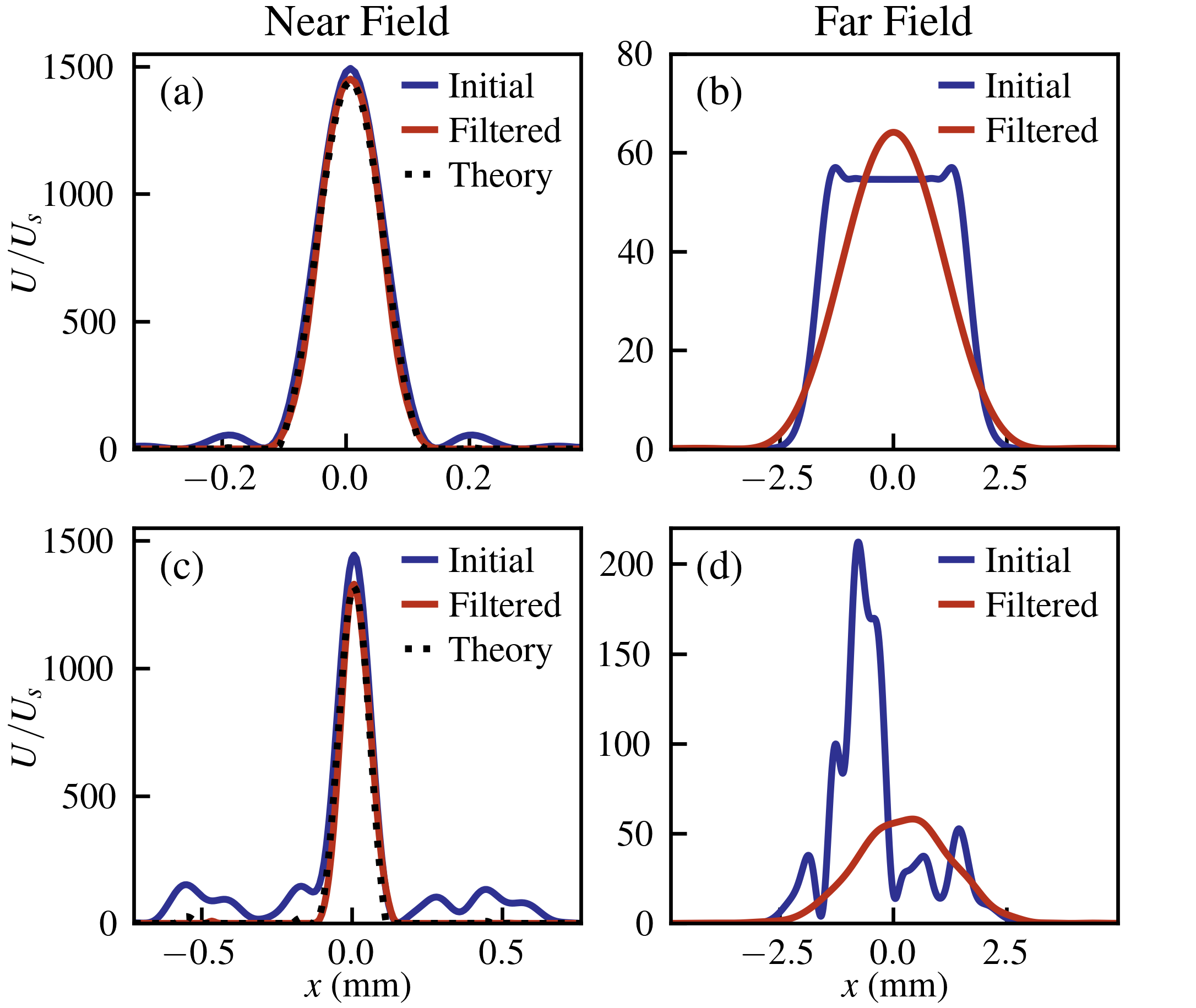}
    \caption{Simulation results: (a) and (b) start with a super-Gaussian beam of order 5 ($r \approx 4\%$) and the triple product is $\beta = 50$; (c) and (d) apply an additional noise on top of the previous profile ($r \approx 10\%$) and the triple product is $\beta = 125$. For both beams, the divergence is $1\,\mathrm{mrad}$ and the peak fluence is $u_0 = 1500$ at the focal plane. The left panels show the beam profiles at the entrance and exit of the gas pinhole. The dotted lines are predictions based on \eqref{eq:eq5}. The right panels show the initial and filtered profiles far from the gas optic.}
    \label{fig:fig_sim}
\end{figure}

The simulation results for two beams with a peak fluence of $u_0 = 1500$ and $r \approx 4\%$ and $r \approx 10\%$, respectively, are shown in Fig.\,\ref{fig:fig_sim}. The first pinhole has a triple product of $\beta = 50$. About $96\%$ of the main beam and $2\%$ of the side lobes are transmitted, corresponding to a noise attenuation of $\Gamma \approx 48$. The second pinhole has a triple product of $\beta = 125$. In this case, about $92\%$ of the main beam and $6\%$ of the side lobes are transmitted, corresponding to $\Gamma \approx 15$. If this were a $266\,\mathrm{nm}$ probe beam cleaned by ozone (i.e. $\sigma \approx 10^{-17}\,\mathrm{cm^2/molecule}$ at $1\,\mathrm{atm}$, $293\,\mathrm{K}$ \cite{daumont1992ozone}, $U_s \approx 75\,\mathrm{mJ/cm^2}$), the triple products would correspond to a $5\%$ ozone mixture and $4\,\mathrm{cm}$ and $10\,\mathrm{cm}$ propagation distances for the two pinholes. In comparison, the Rayleigh length is about $8.5\,\mathrm{cm}$ for $1\,\mathrm{mrad}$ divergence as used in our simulations. 

According to \eqref{eq:eq8} and \eqref{eq:eq9}, the transmission and attenuation would have been $T_0 \approx 97\%$ and $\Gamma \approx 10$ for the first setup and $T_0 \approx 92\%$ and $\Gamma \approx 4$ for the second one. The discrepancy may be due to neglecting diffraction in propagation and approximating energy contained by different modes with a single peak value. Nonetheless, the model is sufficient for suggesting parameters that can produce noticeable spatial cleaning effects. The discrepancy may also diminish as the Rayleigh length becomes much longer than the gas pinhole.

\begin{figure*}[tb]
    \centering
    \includegraphics[width=\linewidth]{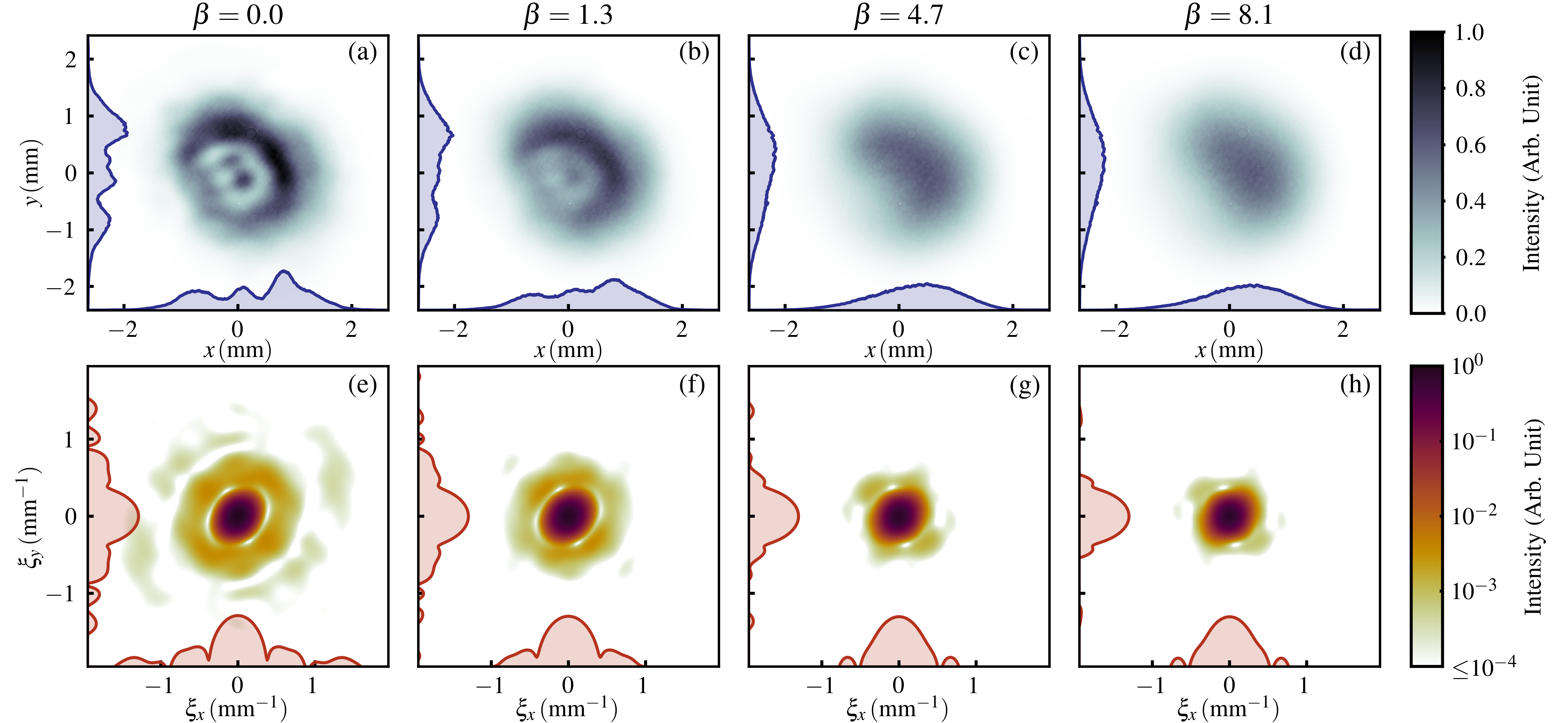}
    \caption{Experimental results. (a)-(d). Spatial profile of the beam cleaned by a gas pinhole at four different triple products, adjusted by changing the ozone concentration. (e)-(f). Fourier spectra of the spatial profiles in the first row in logarithmic scale. The horizontal and vertical shaded curves show the central profile of the images.}
    \label{fig:fig_result}
\end{figure*}

We further demonstrated this process experimentally with the setup in Fig.\,\ref{fig:fig_schematic}. The probe was created by frequency quadrupling a $1064\,\mathrm{nm}$ beam from a Q-switched Nd:YAG laser (Spectra-Physics PIV-400) via temperature-controlled DKDP and KDP crystals. The $266\,\mathrm{nm}$ probe contained an initial energy of $4.2 \pm 0.1\,\mathrm{mJ}$ and had a full width at half maximum (FWHM) pulse duration of $5\,\mathrm{ns}$. The beam was focused into an ozone-oxygen mixture flowing in a $3\,\mathrm{cm}$ wide aluminum tube with a $750\,\mathrm{mm}$ plano-convex lens followed by a $1000\,\mathrm{mm}$ plano-convex lens to achieve a long effective focal length. The diameter of the focal spot was around $0.84\,\mathrm{mm}$ in the gas. The ozone was produced by a corona-discharge-based ozone generator (Oxidation Technologies VMUS-4). The triple product was measured based on the linear absorption of a weak UV probe and reached up to $8.1$ ($\sim1.4\%\,\mathrm{O_3}$). The first row in Fig.\,\ref{fig:fig_result} shows the spatial profiles of the beam with ozone off and at three different ozone concentrations, imaged by sending the beam directly to a CMOS camera (Alvium G5-812 UV). The second row shows the Fourier spectra of the profiles. The high-frequency components were effectively suppressed as we increased the ozone concentration. Based on the Fourier spectra, the initial beam had a noisiness of $r \approx 0.24\%$. The gas pinhole delivered $86\%$ transmission and $1.4$ attenuation at $\beta = 1.3$, $78\%$ transmission and $4.2$ attenuation at $\beta = 4.7$, and $76\%$ transmission and $8.1$ attenuation at $\beta = 8.1$. The performance was likely limited by the extremely poor focal spot quality of the original beam, preventing complete cleaning with a single $3\,\mathrm{cm}$ ozone pinhole.

The experimental results demonstrate that ozone pinholes can significantly improve the spatial quality of an aberrated UV beam. In addition, we expect better temporal contrast since weaker prepulses will also be suppressed. The gas pinhole can be scaled to larger laser systems. For example, at $r = 2\%$ and $u_0 = 10^4$, a gas pinhole with $\beta = ru_0 = 200$ can deliver $t_0 = 98\%$ and $\gamma \approx 280$. The triple product can be achieved using a $5\%$ ozone mixture at $1\,\mathrm{atm}$, $293\,\mathrm{K}$ and an $8\,\mathrm{cm}$ long pinhole. Consider applications involving lasers at $248\,\mathrm{nm}$ or $266\,\mathrm{nm}$ ($U \approx 750\,\mathrm{J/cm^2}$). The normalized fluence corresponds to a $500\,\mathrm{mJ}$ beam with a central waist of $w_0 \approx 0.15\,\mathrm{mm}$, or a $50\,\mathrm{J}$ beam with a central waist of $w_0 \approx 1.5\,\mathrm{mm}$. Scaling to kilojoule- and megajoule-class lasers without exceeding the damage thresholds or using extremely long focal lengths could be achieved with cylindrical lenses or mirrors that focus the beam in one axis and defocus in the other. Spatial modes in different directions will then be cleaned separately.  

Finally, the gas pinhole can be adapted to lasers at different wavelengths, especially those in the visible and ultraviolet regimes. A non-exhaustive list of gases potentially suitable for different wavelengths is summarized in Table \ref{tab:tab1}. In the mid- and far-infrared regime, greenhouse gases including carbon dioxide and water vapor have particularly high absorption cross-sections, although it is unclear whether they will exhibit similar behavior for cleaning pulses. Since the device is insensitive to the pointing of the lasers, a multi-pass spatial filter can be built to achieve a longer effective attenuation distance and allow the use of gases with smaller cross-sections.

\begin{table}[htb]
\centering
\caption{\bf Gases suitable for filtering at different wavelengths.}
\begin{ruledtabular}
\begin{tabular}{cccc}
Gas & $\lambda\,(\mathrm{nm})$ & $10^{20}\times\sigma\,(\,\mathrm{cm^2/molecule})$ & $U_s\,(\mathrm{mJ/cm^2})$\\
\hline
$\mathrm{NH_3}$ & $193$ & $2000$ \cite{cheng2006absorption} & $51$\\
\hline
$\mathrm{O_3}$ & $248$ & $1080$ \cite{burkholder2020chemical} & $74$\\
\hline
$\mathrm{O_3}$ & $266$ & $968$ \cite{burkholder2020chemical} & $77$\\
\hline
$\mathrm{ClO_2}$ & $351$ & $1275$ \cite{burkholder2020chemical} & $44$\\
\hline
$\mathrm{NO_2}$ & $351$ & $47$ \cite{bogumil2003measurements} & $1204$\\
\hline
$\mathrm{NO_2}$ & $532$ & $16$ \cite{bogumil2003measurements} & $2334$\\
\end{tabular}
\end{ruledtabular}
\label{tab:tab1}
\end{table}

In conclusion, we have developed a gas-based spatial filter using saturated absorption in ozone and studied its performance with an analytic model, numerical simulations, and experiments. The results show significantly improved spatial quality of a $266\,\mathrm{nm}$ beam after focusing through a $3\,\mathrm{cm}$ $1.4\%$ ozone-oxygen mixture, suggesting a method for building alignment-insensitive and damage-resistant pinholes for high-repetition-rate high-energy lasers.

\medskip \noindent
{\bf Funding.} This work was partially supported by NNSA Grant DE-NA0004130, NSF Grant PHY-2308641, and the Lawrence Livermore National Laboratory LDRD program (24-ERD-001).

\medskip \noindent
{\bf Acknowledgments.} Lawrence Livermore National Laboratory is operated by Lawrence Livermore National Security, LLC, for the U.S. Department of Energy, National Nuclear Security Administration under Contract DE-AC52-07NA27344.

\medskip \noindent
{\bf Disclosures.} Work at Stanford has been supported in part by Xcimer Energy LLC.

\medskip \noindent
{\bf Data availability.} Data underlying the results presented in this paper are not publicly available at this time but may be obtained from the authors upon reasonable request.

\bibliography{reference}

\end{document}